\documentclass[aps,prc,twocolumn,superscriptaddress,showpacs,showkeys]{revtex4-1}
\usepackage{graphicx}

\begin{document}
\title{Constraints on the axion-electron coupling for solar axions produced by Compton process and bremsstrahlung}
\author{A.V.~Derbin}
\email {derbin@pnpi.spb.ru}
\author{A.S.~Kayunov}
\author{V.V.~Muratova}
\author{D.A.~Semenov}
\author{E.V.~Unzhakov}

\affiliation{St.Petersburg Nuclear Physics Institute, Gatchina, Russia 188300}

\begin{abstract}
The search for solar axions produced by Compton ($\gamma+e^-\rightarrow e^-+A$) and bremsstrahlung-like ($e^-+Z \rightarrow Z+e^-+A$)
processes has been performed. The axion flux in the both cases depends on the axion-electron coupling constant. The resonant excitation of
low-lying nuclear level of $^{169}\rm{Tm}$ was looked for: $A+^{169}$Tm $\rightarrow ^{169}$Tm$^*$ $\rightarrow ^{169}$Tm $+ \gamma$ (8.41
keV). The Si(Li) detector and $^{169}$Tm target installed inside the low-background setup were used to detect 8.41 keV $\gamma$-rays. As a
result, a new model independent restriction on the axion-electron and the axion-nucleon couplings was obtained: $g_{Ae}\times|g^0_{AN}+
g^3_{AN}|\leq 2.1\times10^{-14}$. In model of hadronic axion this restriction corresponds to the upper limit on the axion-electron coupling
and on the axion mass $g_{Ae}\times m_A\leq3.1\times10^{-7}$ eV (90\% c.l.). The limits on axion mass are $m_A\leq$ 105 eV and $m_A\leq$
1.3 keV for DFSZ- and KSVZ-axion models, correspondingly (90\% c.l.).
\end{abstract}
 \maketitle

\section{Introduction}
Axions arise as a result of the solution of the strong CP-problem proposed by Peccei and Quinn \cite{Pec77}. Wienberg \cite{Wei78} and
Wilczek \cite{Wil78} showed that PQ-solution leads to the existence of a neutral spin-zero pseudoscalar particle with a non-zero mass. The
axion mass as well as the strength of axion's coupling with ordinary matter is inversely proportional to the PQ-symmetry breaking scale
$f_A$. The original PQWW-axion model with $f_{A}\approx(\sqrt{2}G_F)^{-1/2}$ has exact predictions for axion coupling with photons,
electrons and nucleons ($g_{A\gamma}$, $g_{Ae}$ and $g_{AN}$ ). The model has been excluded by the series of experiments with radioactive
sources, reactors and accelerators.

Two types of the "invisible" axion models retained the axion in the form required for the solution of CP-violation problem,
while suppressing its interaction with matter. These are the models of "hadronic" or Kim-Shifman-Vainstein-Zakharov (KSVZ) axion
\cite{Kim79,Shi80} and the GUT or Dine-Fischler-Srednicki-Zhitnitskii (DFSZ) axion \cite{Zhi80,Din81}. The models of "invisible"
axion have no certain predictions for $f_A$, and as a result, no restrictions for the axion mass and the coupling constants.
Moreover, the parameters of axion couplings are significantly model dependent.

The hadronic axion does not interact with leptons and ordinary quarks at the tree level, which results in strong suppression of
$g_{Ae}$ constant through radiatively induced coupling \cite{Sre85}. In some models axion-photon coupling may significantly
differ from the original DFSZ or KSVZ couplings by a factor less then $10^{-2}$ \cite{Kap85}. The effective axion-nucleon
coupling depends on ratios of $u$-, $d$- and $s$-quark masses, axial pion-nucleon couplings $F$ and $D$ and poorly constrained
flavor singlet coupling $S$. Moreover, the values of $g_{AN}$  for the DFSZ axion depend on additional unknown parameter
$\cos^2\beta$ which is defined by the ratio of the Higgs vacuum expectation values \cite{Sre85, Kap85}.

The axion mass $m_{A}$ (in $eV$ units) in both models is given in terms of $\pi^0$ properties:
\begin{equation}\label{ma}
m_A=\frac{f_{\pi}m_{\pi}}{f_A}(\frac{z}{(1+z+w)(1+z)})^{1/2}\approx \frac{6.0\times 10^6}{f_A(GeV)}
\end{equation}
where $f_\pi\cong$ 93 MeV is the pion decay constant; $z = m_u/m_d \cong 0.56$ and $w = m_u/m_s \cong0.029$ are quark-mass
ratios. The restrictions on the axion mass appear as a result of the restrictions on the coupling constants $g_{A\gamma}$,
$g_{Ae}$ and $g_{AN}$.

If the axions do exist, then the Sun should be an intense source of these particles. Axions can be efficiently produced in the Sun by the
inverse Primakoff conversion of the photons in the electromagnetic field of the plasma. The resulting axion flux depends on $g_{A\gamma}^2$
and can be detected by the Primakoff conversion of axions to photons in laboratory magnetic fields \cite{Sik83} - \cite{Ari09} or by the
coherent conversion to photons in the crystal detectors \cite{Avi99}-\cite{Bru08}. The expected count rate of photons depends on the
axion-photon coupling as $g_{A\gamma}^4$. The upper limits on the $g_{A\gamma}$ are ($10^{-10}-10^{-9})\;\rm{GeV}^{-1}$ for axions with
mass less than 0.1 eV.

There are two other possible mechanisms of axion production in the Sun: the reactions of solar cycle and the excitation of the
low-lying energy levels of some nuclei by the high solar temperature. The attempts to observe 
the quasi-monochromatic axions emitted in nuclear magnetic transitions were performed in \cite{Mor95}-\cite{Der10}. The reactions of the
resonant excitation of nuclear levels, the axion-to-photon conversion and the axioelectric effect have been used for detection.

In this letter we present the results of the search for axions  emitted from Sun by the Compton process $\gamma+e^- \rightarrow e^- + A$
and by bremsstrahlung $e^-+Z\rightarrow e^-+Z+A$ in the hot solar plasma. The cross sections of the both reactions depend on the
axion-electron coupling constant $g_{Ae}^2$. The axions can be detected in the reaction of the resonant absorption by $^{169}\rm{Tm}$
nuclear target \cite{Der09}. The 8.41 keV $\gamma$-rays and conversion electrons produced by the de-excitation of the first nuclear level
(Fig.1) can be registered. The detection probability of the axions is determined by the product $g_{Ae}^2\times g_{AN}^2$ which is
preferable for small $g_{Ae}$ or $g_{A\gamma}$ values.

The results of laboratory searches for the axion as well as the astrophysical and cosmological axion bounds can be found in
\cite{PDG08,Raf06}. Constrains obtained with the solar axions remain of interest, even if they are less restrictive than astrophysical
arguments, because they are more comparable to the laboratory experiments.

\section{The axion spectrum and the rate of solar axions absorption by $^{169}$Tm nucleus}
If the axions or other axion-like pseudoscalar particles couple with electrons then they are emitted from Sun by the Compton process and by
bremsstrahlung \cite{Kat75}-\cite{Raf86}. The expected spectrum of axions is calculated using theoretical predictions for the Compton cross
section given in \cite{Pos08,Gon08} and the axion bremsstrahlung due to electron-nucleus collisions given in \cite{Zhi79}. The axion flux
is determined for radial distribution of the temperature $T(r)$, density of electrons $N_e(r)$ and nuclei $N_{Z,A}(r)$ given by BS05(OP)
Standard Solar Model \cite{Bah05} based on high-Z abundances \cite{Asp06}.

The original photon flux for the Compton process is taken in accordance with Planck's law of black-body radiation:
\begin{equation}\label{Plank}
    \frac{dN_\gamma}{dE_\gamma}= \frac{8\pi h^{-3}c^{-2} E_\gamma^2}{e^{E_\gamma/kT}-1}=
    \frac{3.95\times10^{32}E_\gamma^2}{e^{E_\gamma/kT}-1},\rm{keV^{-1}cm^{-2}s^{-1}}
\end{equation}

 The axion spectrum is found by integrating photon spectrum $dN_\gamma/dE_\gamma$ and the Compton cross section $d\sigma^c(E_\gamma,E_A,m_A)/dE_A$ over the radial distribution of $T$ and $N_e$:
 \begin{equation}\label{ComptonFlux}
\frac{d\Phi_A}{dE_A}(E_A)= \frac{1}{R_\odot^2}\int\limits_0^{R_\odot} \int\limits_{E_A}^{\infty}
\frac{dN_\gamma}{dE_\gamma}\frac{d\sigma^c}{dE_A}dE_\gamma N_e(r) r^2 dr
\end{equation}

Since $kT\ll m_e$ we use the non-relativistic expression for $\sigma(E_\gamma)$ given in \cite{Pos08,Gon08}. In this case there is a strong
relation between axion and photon energy $E_A \cong E_\gamma$ and we omit the integration over $E_\gamma$.  The obtained spectrum at the
Earth for $g_{Ae}=10^{-11}$ and $m_A=0$ is shown in Fig.{\ref{fig1}, line 1. In the ($1-10$) keV energy range the spectrum is parameterized
with ~1\% accuracy by the expression:
\begin{equation}\label{diffluxCompton}
    \frac{d\Phi_A}{dE_A}= g^2_{Ae} \times 1.33\times10^{33} E_A^{2.98} e^{-0.774E_A},
\end{equation}
where the value of the  flux is given in ${\rm{cm}^{-2}} {\rm{s}^{-1}} {\rm{keV}^{-1}}$ units and the value of the axion energy $E_A$ is
given in keV units. The corresponding solar axion luminosity is calculated to be $L_A =  g^2_{Ae} \times 1.29\times10^{20} L_{\odot}$,
where $L_\odot$ is the solar photon luminosity. This value is in a good agreement with results obtained in \cite{Kra84,Gon08}.

\begin{figure}
\includegraphics[bb = 30 90 500 760, width=8cm,height=10cm]{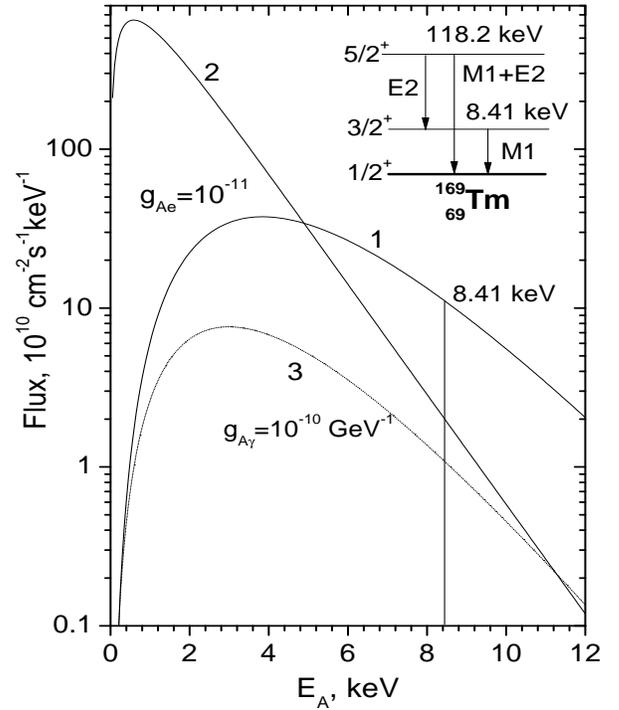}
\caption {1,2 - the spectra of the axions produced by the Compton process and the bremsstrahlung, correspondingly ($g_{Ae}=10^{-11}$,
$m_A=0$). 3 - spectrum of the axions produced by Primakoff effect ($g_{A\gamma}=10^{-10}\rm{GeV}^{-1}$). The level scheme of $^{169}$Tm
nucleus is shown in the inset.}\label{fig1}
\end{figure}

The spectrum of bremsstrahlung axions is calculated in the same way. The Maxwell–-Boltzmann distribution for the kinetic energy of the
scattering electrons at the temperature $T(r)$ is used:

\begin{equation}\label{Maxwell}
    \frac{dN_e}{dE_e}= N_e(r)\frac{2\sqrt{E}e^{-E_e/kT}}{\sqrt{\pi}(kT)^{3/2}}
\end{equation}

The differential cross section for the axion bremsstrahlung process due to electron-nucleus collisions ${d\sigma^b}/{dE_A}(E_e,Z)$ was
calculated in \cite{Zhi79}. The cross section has complex form and it is proportional to $Z^2$. In accordance with \cite{Raf86,Kek08} the
contributions of electron-electron collisions to the axion bremsstrahlung emission are negligible at the some keV's axion energy.
Therefore, only the number density of H, $^4\rm{He}$, $^3\rm{He}$, $^{12}\rm{C}$, $^{14}\rm{N}$, $^{16}\rm{O}$ and $^{26}\rm{Fe}$ nuclei in
a given spherical shell of the solar interior at the radius $r$ are taken into account:

 \begin{equation}\label{BremsFlux}
\frac{d\Phi_A}{dE_A}=  \frac{1}{R_\odot^2}\int\limits_0^{R_\odot} \int\limits_{E_A}^{\infty} \frac{dN_e}{dE_e}\upsilon_e
\frac{d\sigma^b}{dE_A}dE_e \sum\limits_{Z,A}^{} Z^2 N_{Z,A} r^2 dr
\end{equation}

where is cross section for the axion bremsstrahlung due to electron-nucleus collisions given in \cite{Zhi79}.The spectrum of solar
bremsstrahlung axions calculated in assumption that $g_{Ae}=10^{-11}$ and $m_A=0$ is given in Fig.\ref{fig1}, line 2. The spectrum is
softer then the Compton axion spectrum, the maximum axion intensity corresponds to the 0.6 keV energy and the average energy of axions is
1.6 keV. The axion flux is well parameterized by the following expression (in ${\rm{cm}^{-2}} {\rm{s}^{-1}} {\rm{keV}^{-1}}$ units):

\begin{equation}\label{difflux}
    \frac{d\Phi_A}{dE_A}= g^2_{Ae} \times 4.14\times10^{35}  E_A^{0.89} e^{-0.7E_A-1.26\sqrt{E_A}},
\end{equation}

Due to the Compton process and bremsstrahlung the axion fluxes at 8.41 keV are $d\Phi_A/dE_A = g_{Ae}^2\times1.13\times10^{33}
\rm{cm}^{-2}\rm{s}^{-1}\rm{keV}^{-1}$ and $d\Phi_A/dE_A = g_{Ae}^2\times2.08\times10^{32} \rm{cm}^{-2}\rm{s}^{-1}\rm{keV}^{-1}$,
correspondingly. For comparison, the spectrum of axions produced by the Primakoff conversion of photons in electromagnetic field of the
plasma is shown for $g_{A\gamma} = 10^{-10} {\rm{GeV}^{-1}}$ (Fig.\ref{fig1}, line 3).

To take the dependence of axion spectra vs axion mass  into account, the spectra were calculated for different values of $m_A$. The
expected flux of 8.4 keV axions weakly depends on the possible values of $m_A$ for $m_A \leq$ 5 keV, e.g. at $m_A$ = 5 keV the axion flux
decreases by 30 \%.

As a pseudoscalar particle, the axion should be subject to resonant absorption and emission in the nuclear transitions of a magnetic type.
In our experiment we have chosen the $^{169}$Tm nucleus as a target. The energy of the first nuclear level (3/2$^+$) is equal to 8.41 keV,
the total axion flux at this energy is $g_{Ae}^2\times1.34\times10^{33} \rm{cm}^{-2}\rm{s}^{-1}\rm{keV}^{-1}$. The 8.41 keV nuclear level
discharges through $M1$-type transition with $E2$-transition admixture value of $\delta^2$=0.11\% and the relative probability of
$\gamma$-ray emission is $\eta=3.79\times10^{-3}$ \cite{Bag08}.

The cross-section for the resonant absorption of the axions with energy $E_A$ is given by the expression that is similar to the one for
$\gamma$-ray resonant absorption, but the ratio of the nuclear transition probability with the emission of an axion $(\omega_{A})$ to the
probability of magnetic type transition $(\omega_{\gamma})$ has to be taken into account. The rate of solar axion absorption by $^{169}$Tm
nucleus will be
\begin{equation}\label{rate}
    R_A = \pi\sigma_{0\gamma}\Gamma
    \frac{d\Phi_A}{dE_A}(E_A=8.4)\left(\frac{\omega_A}{\omega_\gamma}\right),
\end{equation}
where  $\sigma_{0\gamma}$ is a maximum cross-section of $\gamma$-ray absorption. The experimentally derived value of
$\sigma_{0\gamma}$ for $^{169}$Tm nucleus is $\sigma_{0\gamma}=2.56\times10^{-19}$ cm$^{2}$ \cite{KayLaby}. A lifetime of the
$^{169}$Tm first excited level is $\tau=5.89$ ns \cite{Bag08}, thus the width of energy level $\Gamma=1.13\times10^{-10}$ keV.

The $\omega_A/\omega_\gamma$ ratio calculated in the long-wave approximation, has the following view \cite{Don78,Avi88}:
\begin{equation}\label{omegaaomegag}
    \frac{\omega_A}{\omega_\gamma}=\frac{1}{2\pi\alpha}\frac{1}{1+\delta^{2}}\left[\frac{g^0_{AN}\beta+g^3_{AN}}{(\mu_0-0.5)\beta+\mu_3-\eta}\right]^2\left(\frac{p_A}{p_\gamma}\right)^3.
\end{equation}
Here, $p_\gamma$ and $p_A$ are the photon and axion momenta, respectively, $\mu_0=\mu_p+\mu_n\approx0.88$ and
$\mu_3=\mu_p-\mu_n\approx4.71$ are isoscalar and isovector nuclear magnetic momenta, $\beta$ and $\eta$ are parameters depending on the
particular nuclear matrix elements \cite{Avi88,Hax91}. In case of the $^{169}$Tm nucleus, which has odd number of nucleons and unpaired
proton, in the one-particle approximation the values of $\beta$ and $\eta$ can be estimated as $\beta\approx1.0$ and $\eta\approx0.5$. For
the given parameters the branching ratio can be rewritten as:
\begin{equation} \label{rat}
\frac{\omega_{A}}{\omega_{\gamma}}=1.03(g_{AN}^{0}+g_{AN}^{3})^2(p_A/p_{\gamma})^3.
\end{equation}

In the KSVZ axion model the dimensionless isoscalar and isovector coupling constants $g_{AN}^{0}$ and $g_{AN}^{3}$  are related to $f_A$ by
expressions \cite{Kap85,Sre85}:
\begin{equation}\label{g0g3}
g_{AN}^{0}=-\frac{m_N}{6f_A}[2S+(3F-D)\frac{1+z-2w}{1+z+w}]
\end{equation}
and
\begin{equation}\label{g0g3_2}
g_{AN}^{3}=-\frac{m_N}{2f_A}[(D+F)\frac{1-z}{1+z+w}]
\end{equation}
where $M_N\approx939$ MeV is the nucleon mass. Axial-coupling parameters $F$ and $D$ are obtained from hyperon semi-leptonic decays with
high precision: $F$=0.462 $\pm$ 0.011, $D$= 0.808 $\pm$ 0.006 \cite{Mat05}. The parameter $S$ characterizing the flavor singlet coupling
still remains a poorly constrained one. The boundaries $(0.37\leq S\leq0.53)$ and $(0.15\leq S\leq0.5)$ were found in \cite{Alt97} and
\cite{Ada97}, accordingly.  As a result the value of the sum ($g_{AN}^0+g_{AN}^3$) is determined within a factor of two,  but the ratio
$\omega_A/\omega_{\gamma}$ does not vanish for any value of parameter $S$. With $S=0.5$ the numerical values of axion-nucleon couplings
are: $g_{AN}^{0}= -4.03\times10^{-8}m_A$ and $g_{AN}^{3}= -2.75\times10^{-8}m_A$, $m_A$ is given in eV units.

The values of $g_{AN}^{0}$ and $g_{AN}^{3}$ for the DFSZ axion depend on the PQ charges of the $u$ and $d$ quarks \cite{Kap85,Sre85}. The
charges $X_u$ and $X_d$ have positive-definite values constrained by relation $X_u$ + $X_d$ = 1. In case of $^{169}$Tm $M1$-transition the
value of $(\omega_{A}/\omega_{\gamma})^{DFSZ}$ ratio lies within the interval $\sim$ (0.11$\div$1.74)$(\omega_{A}/\omega_{\gamma})^{KSVZ}$
($S=0.5$). The lower and upper bounds of this interval are defined by values $X_u=0,X_d=1$ and $X_u=1,X_d=0$  respectively.

In accordance with (\ref{difflux})-(\ref{rat}), the rate of axion absorption by $^{169}$Tm nucleus (\ref{rate}) dependent only on the
coupling constants is (the model-independent view):
\begin{equation}\label{rategamag0g3}
    R_A=1.55\times10^5 g_{Ae}^2(g_{AN}^{0}+g_{AN}^{3})^2(p_A/p_\gamma)^3, \rm{s}^{-1}.
\end{equation}
Using the relations between $g_{AN}^0$, $g_{AN}^3$ and axion mass given by KSVZ model (\ref{g0g3}), the absorption rate can be
presented as a function of $g_{Ae}$ and axion mass $m_A$:
\begin{equation}\label{rategama}
    R_A=5.79\times10^{-10}g_{Ae}^2m_A^2(p_A/p_\gamma)^3,  \rm{s}^{-1}.
\end{equation}

In DFSZ-axion models the parameter $g_{Ae}$ is associated with mass of the electron $m$, so that
\begin{equation}\label{GaeeDFSZ}
 g_{Ae}=(1/3){\cos^2\beta} m/f_{A},
 \end{equation}
where $\beta$ is an arbitrary angle. If one sets $\cos^2\beta$=1 the axion-electron coupling is related to axion mass as
$g_{Ae}$=2.8$\times10^{-11}m_A$, where $m_A$ is expressed in eV units.

The hadronic axion has no tree-level coupling to the electron, but there is an induced axion-electron coupling at the one-loop level
\cite{Sre85}:

\begin{equation}
g_{Ae}=\frac{3\alpha^{2}Nm}{2\pi
f_{a}}\left(\frac{E}{N}\ln\frac{f_{A}}{m}-\frac{2}{3}\frac{4+z+w}{1+z+w}\ln\frac{\Lambda}{m}\right)\label{Gaee}
\end{equation}
where $N$ and $E$ are the model dependent coefficients of the electromagnetic and color anomalies, $\Lambda\approx$ 1 GeV is the cutoff at
the QCD confinement scale. The numerical value of $g_{Ae}$ for $E/N$ = 8/3, which is characteristic for GUT models, and for $N$ = 3 is
$g_{Ae}=6.6\times10^{-15}\left(\frac{8}{3}\ln\left(\frac{1.2\times10^{10}}{m_{A}}\right)-14.6\right)m_{A}$, where $m_{A}$ is expressed in
eV units.  The interaction strength of the hadronic axion with the electron is suppressed by a factor $\sim\alpha^{2}$. Because the
coupling of DFSZ and KSVZ axions with electrons is much weaker than with nucleons the search for effect proportional to $g_{AN}\times
g_{Ae}$ is preferable. The dependence of axion absorption rate vs axion mass can be found through the relations (\ref{GaeeDFSZ}) and
(\ref{Gaee}).

The amount of observed $\gamma$-rays that follow the axion absorption depends on the number of target nuclei, measurement time
and detector efficiency, while the probability of 8.4 keV peak observation is determined by the background level of the
experimental setup.

\section{Experimental setup}
To search for quanta with an energy of 8.41keV, the planar Si(Li) detector with a sensitive area diameter of 66 mm and a thickness of 5 mm
was used. The detector was mounted on 5 cm thick copper plate that protected the detector from the external radioactivity. The detector and
the holder were placed in a vacuum cryostat and cooled to liquid nitrogen temperatures. A $\rm{Tm}_2\rm{O}_3$ target of 2 g mass was
uniformly deposited on a plexiglas substrate 70 mm in diameter at a distance of 1.5 mm from the detector surface. External passive
shielding composed of copper, iron and lead layers was adjusted to the cryostat and eliminated external radioactivity background by a
factor of about 500.

The experimental setup was located on the ground surface. Events produced by cosmic rays and fast neutrons were registered by an active
shielding consisting of five plastic scintillators $50\times50\times12$ cm in size. The rate of 50 $\mu$s veto signals was 600 counts/s,
that lead to $\approx$ 3\% dead time.  The Si(Li) detector was sectionalized into nine separate sections in order to lower the capacities
of individual detectors and subsequently increase the resulting energy resolution. Every section was equipped with a charge-sensitive
preamplifier with resistive feedback, a shaping amplifier, and a 12-step analog-to-digital converter. 18 spectra consisting of 4096
channels from each detector (in anti- and in coincidence with the veto signals) are obtained.

Though the detector amplifications were virtually the same, the energy calibrations were performed for each detector independently.
Standard calibration sources $^{57}\rm{Co}$ and $^{241}\rm{Am}$ were used. Energy resolution in the integral spectrum for a 14.4 keV
$\gamma$-ray line was $\sigma$ = 0.63 keV. The high energy resolution and accurate knowledge of the energy scale are crucial to our
experiment because the energies of the characteristic X-rays of thulium are close to 8.41 keV. The most intense L-lines in the case of
vacancy on K-shell have the next energies and intensities: 7.18 keV (8.1\%, L$_{\alpha 1}$), 8.10 keV (5.2\%, L$_{\beta 1}$) and 8.47 keV
(1.6 \%, L$_{\beta 2}$).

The sensitive volume and the area of the Si(Li) detector were measured using the X-ray and $\gamma$-lines of a standard $^{241}\rm{Am}$
source. The detection efficiency for an energy of 8.41 keV was estimated by numerical M-C simulation, taking into account the
self-absorption of $\gamma$-rays by the target. The simulation results were checked on a $^{241}\rm{Am}$ source placed behind the
Tm$_2$O$_3$ target. The total $\gamma$-ray detection efficiency at an energy of 8.41 keV was $\varepsilon = (6.16 \pm  0.30) \%$.

\section{Results}
Measurements were made over 31.8 days of live time by two hour series to monitor the stability of the Si(Li)- and the scintillation
detectors. The integral energy spectrum of Si(Li)-detectors measured in the anticoincidence with signals of the veto scintillators is shown
on the inset in Fig.\ref{fig2}. One can clearly identify peaks related to X-rays of thulium ($\rm{K}_{\alpha1}$ = 50.74 keV and
$\rm{K}_{\alpha2}$ = 49.77 keV). There were no statistically significant peaks in the spectrum of events correlated with the veto signals.

\begin{figure}
\includegraphics[bb = 30 90 500 760, width=8cm,height=10cm]{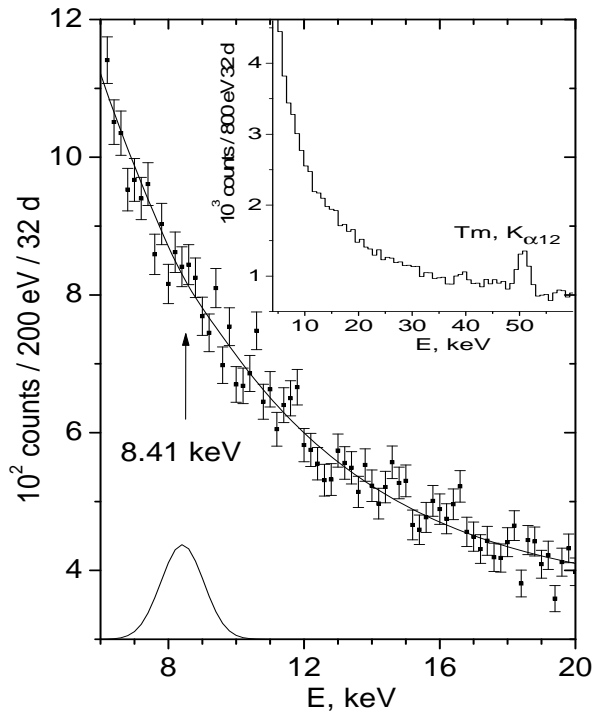} \caption { The Si(Li)-detector energy spectra
measured in the anticoincidence with the veto signal. Solid line shows the fitting result in the 6 - 20 keV range corresponding to the
minimum $\chi^2$. The spectrum in the ($4-60$) keV region is shown in inset.}\label{fig2}
\end{figure}

Fig. 2 shows the detailed energy spectrum within the (6$\div$20) keV interval, where the 8.41 keV is expected. The maximum likelihood
method was used to define the intensity of the 'axion peak'. As the likelihood function $N(E)$, we took the sum of the exponential function
for a smooth background and a Gaussian:
\begin{equation}\label{likelihood}
N(E)=a+b\times\exp(cE)+\frac{S_A}{\sqrt{2\pi}\sigma}\exp\left[-\frac{(E_{0}-E)^{2}}{2\sigma^{2}}\right].
\end{equation}
The energy resolution $\sigma$ and peak position $E_0$ were fixed, while the intensity $S_A$ and background parameters $a$, $b$, and $c$
were free during the fitting. The result of the fit corresponding to the minimum of $\chi^2$ is shown in Fig. 2 by solid line. A standard
method was used to set the upper limit on the 8.41 keV peak intensity: $\chi^2$ was determined for different fixed values of $S_A$ while
all other parameters were free. Obtained probability function $P(\chi^2(S))$ was normalized to unity for $S\geq 0$. The upper limit
estimated in this manner was $S_{lim}$ = 217 at a 90\% confidence level.

The expected number of registered 8.41 keV $\gamma$-quanta for the detection rate defined by (\ref{rategamag0g3}) is
\begin{equation}\label{slim}
    S_A=\varepsilon \eta N_{169Tm} T R_A = 4.0\times 10^{24} R_A \leq S_{lim},
\end{equation}
where  $N_{169Tm}=6.23\times10^{21}$ - the number of $^{169}$Tm nuclei, $T=2.75\times10^{6}$ s - time of measurement,
$\varepsilon=6.16\times10^{-2}$ - detection efficiency and $\eta=3.79\times10^{-3}$ - internal conversion ratio \cite{Bag08}.

The upper limit on axions absorption rate by $^{169}\rm{Tm}$ nucleus $R_A \leq 5.43\times10^{-23}  \rm{s}^{-1}$ set by our experiment
limits the possible values of coupling constants $g_{Ae}$, $g_{AN}$ and axion mass $m_A$. According to (\ref{rategamag0g3}) and
(\ref{rategama}) and taking into account the approximate equality of the axion and $\gamma$-quantum momenta $(p_A/p_{\gamma})^3 \simeq 1$
for $m_A\leq2$ keV we obtain (at 90\% c.l.):
\begin{equation}\label{lim1}
g_{Ae}\times |(g_{AN}^0+g_{AN}^3)| \leq 2.1\times 10^{-14}
\end{equation}
\begin{equation}\label{lim2}
g_{Ae}\times m_A \leq 3.1\times 10^{-7}\;\rm{eV}
\end{equation}
The restriction (\ref{lim1}) is a model independent one on axion (or any other pseudoscalar particle) couplings with electron and nucleons.
The result (\ref{lim2}) presented as a restriction on the range of possible values of $g_{Ae}$ and $m_A$ (the relations (\ref{g0g3}) and
(\ref{g0g3_2}) between $g_{AN}$ and $m_A$ are used) allows one to compare our result with results of other experiments restricting $g_{Ae}$
(Fig.\ref{fig3}). The limits on $g_{Ae}\times m_A$ for DFSZ axion lie in the range ($0.33-1.32$) of the restriction (\ref{lim2}).

The relation (\ref{lim2}) excludes the region of relatively large values of $g_{Ae}$ and $m_A$ possible in KSVZ and DFSZ models. The
strongest limit on $g_{Ae}\simeq 9\times10^{-11}$ corresponds to $m_A\simeq$ 5 keV. The hypothesis of keV-scale bosons as possible dark
matter candidates has been considered in \cite{Ber06, Pos08}. More stringent limits on $g_{Ae}$ was found under assumption that axion
luminosity does not exceeds 0.1 $L_{\odot}$ \cite{Gon08}. The recent constraints on keV-mass pseudoscalar dark matter by CoGeNT
\cite{Aal08} and CDMS \cite{Ahm09} are more restrictive than the solar luminosity limit. Reactor, beam dump and positronium decay
experiments constrain the MeV-scale region of  axion masses \cite{Alt95,Cha07,Bor08,Kon86,Bjo88,Asa91}. Very strong restriction on $g_{Ae}$
can be obtained for axions with mass in ($1.1-5.4$) MeV range from the positron flux near the Earth's atmosphere surface \cite{Der10}.

\begin{figure}
\includegraphics[bb = 30 90 500 760, width=8cm,height=10cm]{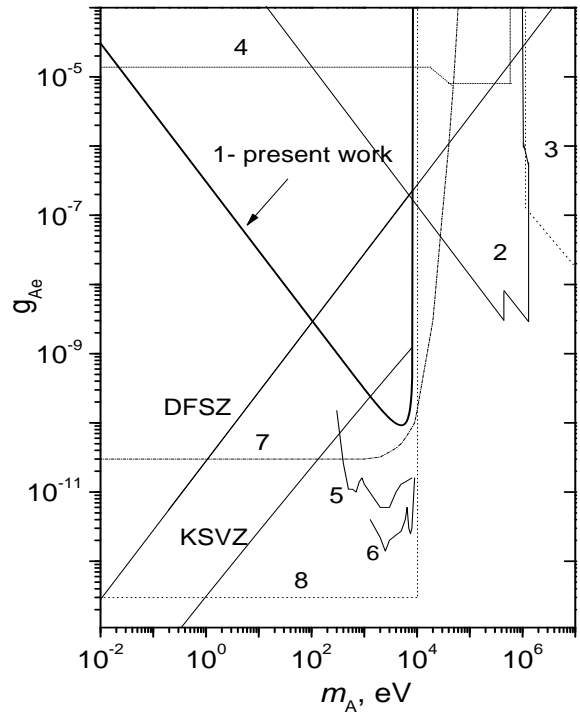}
\caption {The limits on $g_{Ae}$ coupling constant obtained by 1- present work,  2- reactor experiments and Borexino
\cite{Alt95,Cha07,Bor08}, 3- beam dump experiments \cite{Kon86,Bjo88},  4- ortopositronium decay \cite{Asa91}, 5 - CoGeNT \cite{Aal08}, 6-
CDMS \cite{Ahm09}, 7 - solar axion luminosity \cite{Gon08} 8 - red giant \cite{Raf06}). The areas of excluded values are located above the
corresponding curves. The expected values of $g_{Ae}$ and $m_A$ in DFSZ and KSVZ axion models are also shown. } \label{fig3}
\end{figure}

The obtained constraints (\ref{lim1}) and (\ref{lim2}) are valid in assumption that axions escape without restraint from the
Sun. Axions leaving the center of the Sun pass through the matter layer with $\approx 6.8\times 10^{35}$ electrons/cm$^2$. As
the result the absorption of the axions due to inverse Compton process $A+e^-\rightarrow e^-+\gamma$ is lower than 10$\%$, if
$g_{Ae}\leq 1\times$10$^{-5}$. The other process associated with axion-electron coupling is the axioelectric effect
$A+e+Z\rightarrow e+Z$. The cross section of axioelectric effect was calculated in \cite{Zhi79,Pos08}. The cross section has a
$Z^{5}$ dependence and solar abundance of Fe constrains the sensitivity of experiment by value $g_{Ae} < 8\times 10^{-6}$.

Using the limit (\ref{lim1}) and relations (\ref{Gaee}) and (\ref{GaeeDFSZ}) one can obtain the limit on the mass of KSVZ axion - $m_A \leq
1.3$ keV and DFSZ axion  $m_A \leq 105$ eV ($\cos^2\beta$=1)  (90\%c.l.).

A search for solar axions by resonant absorption or through axioelectric effect was made in \cite{Mor95}-\cite{Der10}. The
strongest limit for the hadronic axion mass ($m_A \leq$ 151 eV) was made for 14.4 keV axions emitted in the M1 transition of a
$^{57}\rm{Fe}$ nucleus \cite{Der09A}. A significant advantage of our experiment is that for the M1 transition of $^{169}\rm{Tm}$
(as opposed to $^{57}\rm{Fe}$), $\omega_A/\omega_\gamma$ ratio depends weakly on the actual values of $S$ and $z$.

In our case, the uncertainty in $S$ allows the upper limit to change from $g_{Ae}\times m_A \leq2.7\times10^{-7}$ eV ($S$ = 0.7) to
$g_{Ae}\times m_A \leq3.6\times 10^{-7}$ eV ($S$ = 0.3)(Fig.\ref{fig4}). The value of $u$- and $d$-quark-mass ratio $z$ = 0.56 is generally
accepted for axion papers, but it could vary in the range ($0.35-0.6$) \cite{PDG08}. This uncertainty changes the obtained constraints
insignificantly: from $g_{Ae}\times m_A \leq2.4\times10^{-7}$ eV ($z$ = 0.35) to $g_{Ae}\times m_A \leq3.2\times10^{-7}$ eV ($z$ = 0.6).

\begin{figure}
\includegraphics[bb = 30 90 500 760, width=8cm,height=10cm]{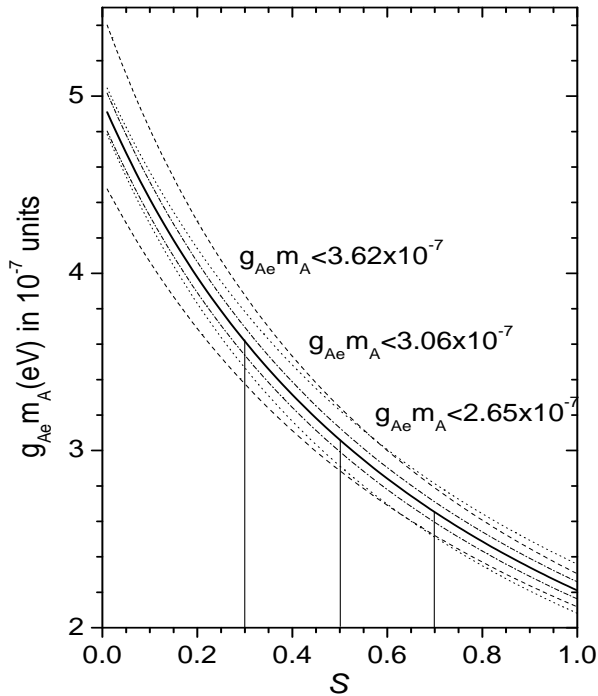}
\caption { The limit on $g_{Ae} \times m_A$ (in $10^{-7}$ eV units) versus the value of $S$ parameter ($\beta$=1, $\eta$=0.5, $z$=0.56,
solid line). The doted, dash-doted and dashed lines correspond to the values of $\beta$, $\eta$ and $z$ changed by $\pm$10\%,
correspondingly.} \label{fig4}
\end{figure}

The sensitivity of the experiment depends on the total efficiency of registration which is  $\eta \times \varepsilon\approx 2\times
10^{-4}$ in our case. This value can be increased significantly by introducing the Tm target inside the sensitive volume of detector.

\section{Conclusion}
We searched for the resonant excitation of the first nuclear level of $^{169}\rm{Tm}$ (8.41 keV) by axions formed inside the Sun
due to Compton and bremsstrahlung process provided by axion-electron coupling. A sectionalized Si(Li) detector installed inside
a low background setup was used to register 8.41 keV $\gamma$-quanta. As a result,  we obtained a new model independent upper
limit on the axion-electron and axion-nucleon couplings: $g_{Ae}\times |g^0_{AN}+ g^3_{AN}|\leq 2.1\times 10^{-14}$ (90\% c.l.).
In model of hadronic axion this restriction corresponds to the upper limit on the axion-electron coupling and on axion mass
$g_{Ae}\times m_A \leq3.1\times10^{-7}$ eV (90\% c.l.).


\end{document}